# Multi-Scale Irregularities Product: a data product utilizing the high-resolution Swarm plasma density data for space weather applications

**Authors**: Yaqi Jin[1], Luca Spogli[2], Daria Kotova[1], Alan Wood[3], Jaroslav Urbář[4], Lucilla Alfonsi[2], Mainul Mohammed Hoque[5], and Wojciech Miloch[1]
(1) Department of Physics, University of Oslo, Oslo, Norway
(2) Istituto Nazionale di Geofisica e Vulcanologia, Rome, Italy
(3) UK Met Office, FitzRoy Road, Exeter, EX1 3PB, UK
(4) Institute of Atmospheric Physics of the Czech Academy of Sciences (IAP), Czech Republic
(5) Deutsches Zentrum für Luft- und Raumfahrt (DLR), Germany

[*]Corresponding author: Yaqi Jin (yaqi.jin@fys.uio.no)

**Abstract**

The Earth's ionospheric dynamics are of multi-scale nature, and the small-scale phenomena have been difficult to measure by in-situ techniques because of low data resolution. In this paper, we make use of the high-resolution Swarm faceplate plasma density data at 16 Hz to develop a set of parameters that can characterize multi-scale ionospheric structures and irregularities along the Swarm orbit. We present the methods for calculating density gradients over different window sizes, rate of change of density index, power spectral density and the spectral slope at both low and high latitudes. The faceplate plasma data are not continuously available through the years. However, about 8 years of data from Swarm A are processed from late 2014 to the end of 2025. Some statistical results from Swarm A are presented. The variations of plasma structures and irregularities are dependent on solar activity, season, local time and geomagnetic activities, and the variations show different patterns between low and high latitudes. For example, the high-latitude ionosphere is characterized by persistent ionospheric structures and irregularities poleward of ±60º magnetic latitude, while the low-latitude ionospheric irregularities are only dominant during 19-01 local time near the magnetic equator. The occurrence of steep spectral slope at high latitudes shows clear seasonal variations, i.e., it maximizes during local summer and minimizes during local winter in both hemispheres. However, the occurrence of steep spectral slope at low latitudes is only sensible when significant plasma structures and irregularities are present. We further calculate the histogram of spectral slopes at low latitudes when the rate of change of density index is enhanced. The histogram resembles a Gaussian distribution with an expected value of 1.97. The processed data are available to the wider community. Given the high resolution, this new data product will be useful for the scientific communities that are interested in the magnetosphere-ionosphere-thermosphere coupling and near-Earth space environment.

1. Introduction

Due to the complex coupling processes with the solar wind, magnetosphere, thermosphere and lower atmosphere, the Earth's ionospheric dynamics are of multi-scale nature (e.g., Moen et al., 2012; Nishimura et al., 2021). For example, ionospheric irregularities exhibit a wide range of spatial scales, from very large scale phenomena such as tongue of ionization (TOI) that extends over a horizontal spatial scale of 1000 km (Foster et al., 2005), polar cap patches of 100-1000 km (Moen et al., 2006), and equatorial plasma bubbles of 1000s km along magnetic field and 100s km in the longitudinal direction (Balan et al., 2018; Bhattacharyya, 2022), to small-scale irregularities of a few meters that scatter radar signals (e.g., Fukao et al., 2004; Ovodenko et al., 2020). Due to the scintillation effect of the Global Navigation Satellite System (GNSS) signals, ionospheric irregularities of ~400 m have gained growing interests since the early 2000s (Kintner et al., 2007). Ionospheric scintillations are rapid fluctuations in the received phase and amplitude of the trans-ionospheric radio signals (Yeh & Liu, 1982). Historically, ionospheric scintillation receivers have been used to investigate the occurrence and global extent of ionospheric irregularities (e.g., Aarons, 1982; Basu et al., 1988). In recent decades, there have been more and more studies using GNSS scintillation receivers owing to the more easily accessible and straightforward use of the GNSS techniques (Mitchell et al., 2005; Kintner et al., 2007; Spogli et al., 2009; Alfonsi et al., 2011; Jin et al., 2014, 2015; Wang et al., 2016). On the other hand, the ground-based instrumentation is still challenging to be deployed in the ocean and remote areas.

Low Earth Orbit (LEO) satellites provide another means to monitor ionospheric irregularities by measuring in-situ densities and they become a valuable source for the global statistics and morphology of ionospheric irregularities (Xiong et al., 2010; Zhang et al., 2017; Zakharenkova & Astafyeva, 2015; Aa et al., 2020; Kotova et al., 2022). Amongst the LEO satellites, the European Space Agency Swarm constellation provides an important source of in-situ observations, as it is equipped with complementary scientific instruments (Friis-Christensen et al., 2006). Despite being conceived primarily as a magnetic mission, Swarm has proven capable of providing ionospheric information of paramount importance (see e.g., Wood et al., 2022 and references therein).

To monitor ionospheric irregularities around the globe, Jin et al. (2019) developed the Ionospheric Plasma IRregularities (IPIR) data product, which contains a number of parameters to characterize ionospheric irregularities at different scales (Jin et al., 2022). The IPIR product is now an official data product for Swarm, and it is available through https://swarm-diss.eo.esa.int/#swarm%2FLevel2daily%2FLatest_baselines%2FIPD. Since the release, the IPIR data has proven to be a useful dataset for studying ionospheric irregularities (Wood et al., 2022; De Michelis et al., 2021; Buschmann et al., 2024; Song et al., 2023; Urbar et al., 2022; Jin & Xiong, 2020; Jin et al., 2025; Wood et al., 2024). For example, Kotova et al. (2023) validated the IPIR product by comparing with data from ground-based GNSS scintillation receivers across polar, auroral, and low-latitude regions, showing that IPIR serves as a viable proxy for identifying the severity of ionospheric fluctuations and the likelihood of trans-ionospheric radio signal scintillation. Furthermore, the IPIR dataset has also been used to develop and evaluate statistical models of the variability of plasma in the topside ionosphere (Wood et al., 2024; Spogli et al., 2024).

At a sampling rate of 2 Hz and an orbital velocity of ~7.5 km/s, the Swarm Langmuir probe (LP) plasma density data is limited by the Nyquist frequency to resolving ionospheric structures with spatial scales of ~7.5 km or greater (Jin et al., 2020). Besides the LP, plasma density can also be derived from the electric current collected by the faceplate (FP) of the Thermal Ion Imager (TII) at 16 Hz. This makes it possible to study plasma structures and irregularities down to sub-kilometer (Ivarsen et al., 2019; Ivarsen et al., 2021a; Aol et al., 2022; Zheng et al., 2025). The FP plasma density data is obtained by assuming that the current is carried by ions hitting the faceplate due to the orbital motion of the spacecraft (Buchert, 2017). The equation for the current follows:

$$I_{fp} = -eAu_iN_i \qquad (1)$$

where $I_{fp}$ is the faceplate current in Ampere, $e$ is the electron charge, $A$ is the faceplate area, $u_i$ is the ion velocity relative to the faceplate surface, and $N_i$ is the ion density. It is assumed that the electron density $N_e$ is equal to the ion density $N_i$ due to quasi-neutrality of the ionospheric plasma. In the following, we will not differentiate electron density, ion density and plasma density. To avoid the modulation of the LP harmonic submode, the plasma density from the FP current is only derived when the bias of TII is below -2.5 V (typical – 3.5 V) (Buchert, 2017).

The Swarm Space Weather: Variability, Irregularities, and Predictive Capabilities for the Dynamic Ionosphere (Swarm-VIP-Dynamic) project, within the Swarm 4D-Ionosphere framework, aims to study ionospheric dynamics and irregularities and to improve predictive capability using advanced dynamic models. A key objective is to incorporate small-scale ionospheric structures and irregularities using FP plasma density data. In this paper we present the processing algorithms for the MUlti-Scale Irregularities produCt (MUSIC) dataset, describe its data availability, and report selected statistical results.

## 2. Processing algorithms

### 2.1 Density spatial gradient $\nabla Ne$

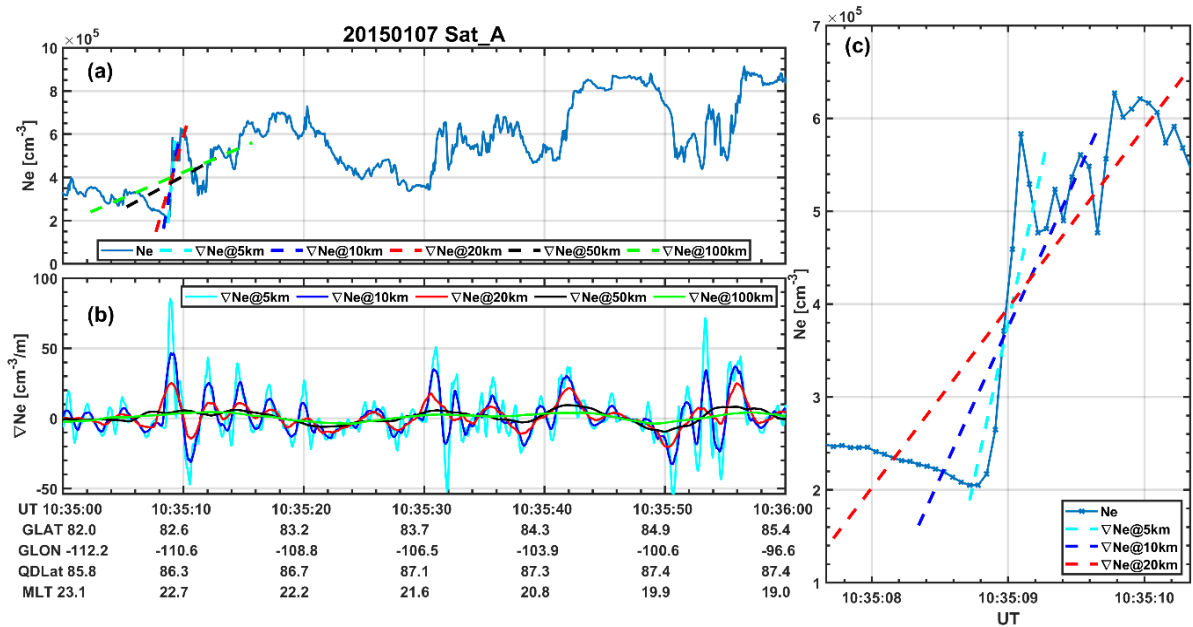

Figure 1. (a) The raw electron density (16 Hz) and density regressions (density gradients) over different window sizes. (b) The density gradients calculated in respective running windows. (c) The expanded view of the calculated density gradients over 5 km, 10 km and 20 km, respectively. UT= Universal Time; GLAT = Geographic Latitude; GLON = Geographic Longitude; QDLat = Quasi-Dipole Latitude; MLT = Magnetic Local Time.

The first category of parameters to characterize horizontal ionospheric irregularities is density gradients at various spatial scales. Since we use 16 Hz data, the smallest scale we can resolve is 469 m, assuming the Swarm satellite speed of 7.5 km/s. We derive the density gradients at 5 km, 10 km, 20 km, 50 km and 100 km using running windows calculated via linear regression over 11, 21, 43, 107, and 213 data points, respectively. Figure 1 gives an example of the raw electron density and the calculated density gradients on January 7, 2015 for Swarm A. For each product the points used for linear regression are centred on the timestamp, for example $\nabla Ne@5km$ uses 5 data points from before and after the timestamp stated. The one minute of data shows observations in the northern polar cap (> 85° quasi-dipole Latitude (QDLat)). Figure 1a shows the original electron density in blue, where polar cap patches are observed as indicated by electron density enhancements. Significant density gradients are associated with these enhancements. We calculate the density gradients using linear regression across various running windows; these are exemplified as dashed lines in Figure 1a. If there is missing data (filled by NaN), the calculated density gradients will be NaN as well. Figure 1c shows an expanded view near 10:35:09 UT. Smaller window sizes are more effective at capturing sharp density gradients, whereas larger window sizes tend to smooth the calculated density gradients. The time series of the density gradients at different spatial scales are shown in Figure 1b. The small-scale density gradients (e.g., $\nabla Ne@5km$) usually show much higher values than the large-scale gradients (e.g., $\nabla Ne@100km$). This suggests that sharp small-scale gradients are often missing when using low-resolution in-situ data (e.g., IPIR dataset (Jin et al., 2022)).

## 2.2 RODI

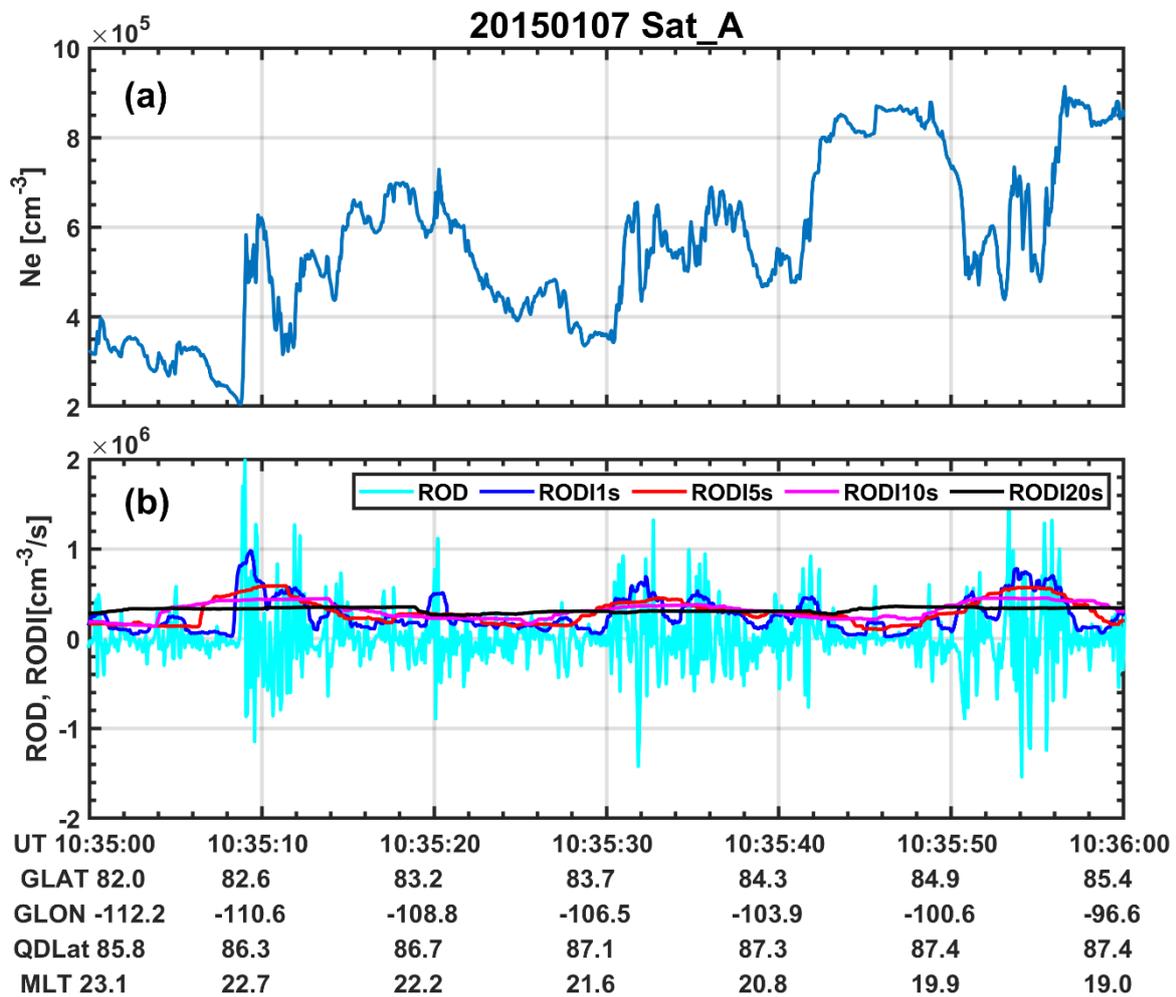

Figure 2. (a) The raw electron density (16 Hz). (b) The density variations characterized by *ROD*, *RODI1s*, *RODI5s*, *RODI10s*, and *RODI20s*, respectively.

We calculate the rate of change of density (*ROD*) and the rate of change density index (*RODI*) as follows:

$$ROD(t) = \frac{Ne(t + \Delta t) - Ne(t)}{\Delta t} \tag{2}$$

where $N_e(t)$ is the electron density at time t, $\Delta t = 0.0625$ s since we use 16 Hz data. This parameter can capture ionospheric irregularities of small spatial scales (~ 469 m). *RODI* is the standard deviation of *ROD* in different running windows:

$$RODI(t) = \sqrt{\frac{1}{N-1} \sum_{t_i=t-\Delta t/2}^{t_i=t+\Delta t/2} \left|ROD(t_i) - \overline{ROD(t)}\right|^2} \tag{3}$$

where $\overline{ROD(t)}$ is the mean value of *ROD* at time *t*:

$$\overline{ROD(t)} = \frac{1}{N} \sqrt{\sum_{t_i=t-\Delta t/2}^{t_i=t+\Delta t/2} ROD(t_i)} \tag{4}$$

The interval $\Delta t = $ 1 s, 5 s, 10 s, and 20 s is used for *RODI1s*, *RODI5s*, *RODI10s*, and *RODI20s*, respectively. Figure 2 presents an example of raw electron density, *ROD*, *RODI1s*, *RODI5s*, *RODI10s*, and *RODI20s* in the same time period of Figure 1.

### 2.3 Spectral slope

Another way to characterize ionospheric irregularities is to investigate the power spectra. The power spectral density (PSD) of ionospheric plasma irregularities can often be expressed by a power law in the form:

$$P(f) \propto f^{-p}, \tag{5}$$

where f is the frequency, $P(f)$ is the PSD, and *p* is the spectral slope or spectral index (Kelley et al., 1982; Tsunoda, 1988). The spectral slope can be used to reveal different instability processes for energy injection and dissipation processes (e.g., Kintner & Seyler, 1985; Ivarsen et al., 2019). Figure 3 presents examples of PSD and spectral slopes in the northern polar cap under quiet conditions (Kp = 1). To calculate the power spectra and spectral slopes, as shown in Figure 3 we have tested several window sizes (8 s, 10 s, 20 s, 30 s, 40 s, and 60 s), which correspond to horizontal spatial scales of 60 km, 75 km, 150 km, 225 km, 300 km, and 450 km, respectively. For example, Figure 3a is obtained by calculating the power spectral density during the interval 10:42:56-10:43:04 UT, i.e., 8 s of data. There are slight differences in the PSD. The PSD at low frequencies is smoother for shorter window sizes. Figure 4 shows the time series of the high-latitude plasma density (Figure 4a) and spectral slopes (Figures 4b-4f). The geographic and magnetic coordinates are shown in the bottom of the figure. In Figure 4b-4f, different colors are used to present spectral slopes calculated using different time steps (resolutions). For example, the blue line presents the spectral slope calculated at a time step of 0.0625 s (16 Hz), while the green line is calculated at a time step of 1 s. The longer time steps can be considered as downsampled from the shorter time steps. For completeness, we also show the case for the equatorial region in Figures 5 and 6. This event was during a geomagnetic storm on September 8, 2017 (Kp = 5- at the time). Plasma irregularities can be very localized. Thus, increasing the window size can smooth out the PSD and lower the spectral slopes with regions without significant irregularities. Therefore, we need to choose a window size and time step based on the consideration of robustness of the spectral slope calculation and the scale size of plasma structures and irregularities. At the end, we decided to adopt the window size of 10 s and time step of 1 s (which corresponds to the green lines in Figures 4c and 6c) as the best tradeoff.

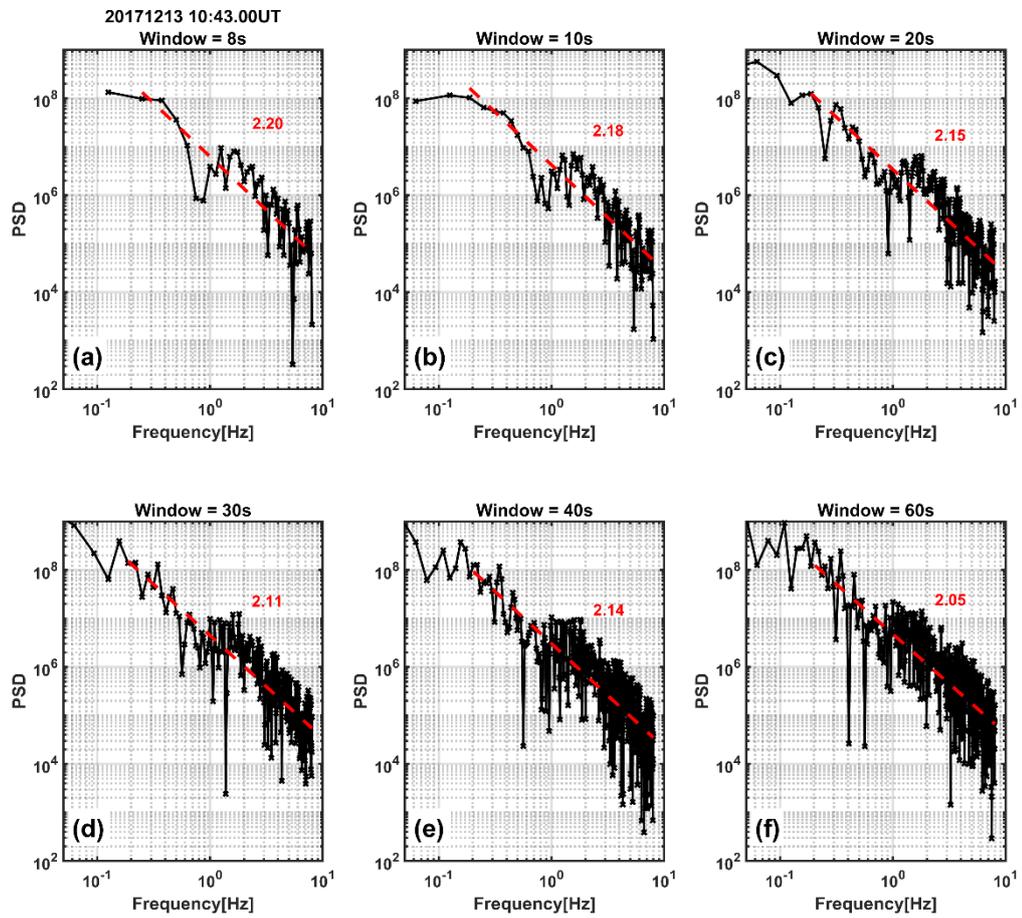

Figure 3. The power spectral density (PSD) and the fitted spectral slopes at frequency range of 0.2-8 Hz. Each panel presents the PSD for a specific window size. The window sizes of 8 s, 10 s, 20 s, 30 s, 40 s, and 60 s correspond to horizontal spatial scales of 60 km, 75 km, 150 km, 225 km, 300 km, and 450 km, respectively. The spectral slopes are annotated by the red dashed lines and numbers.

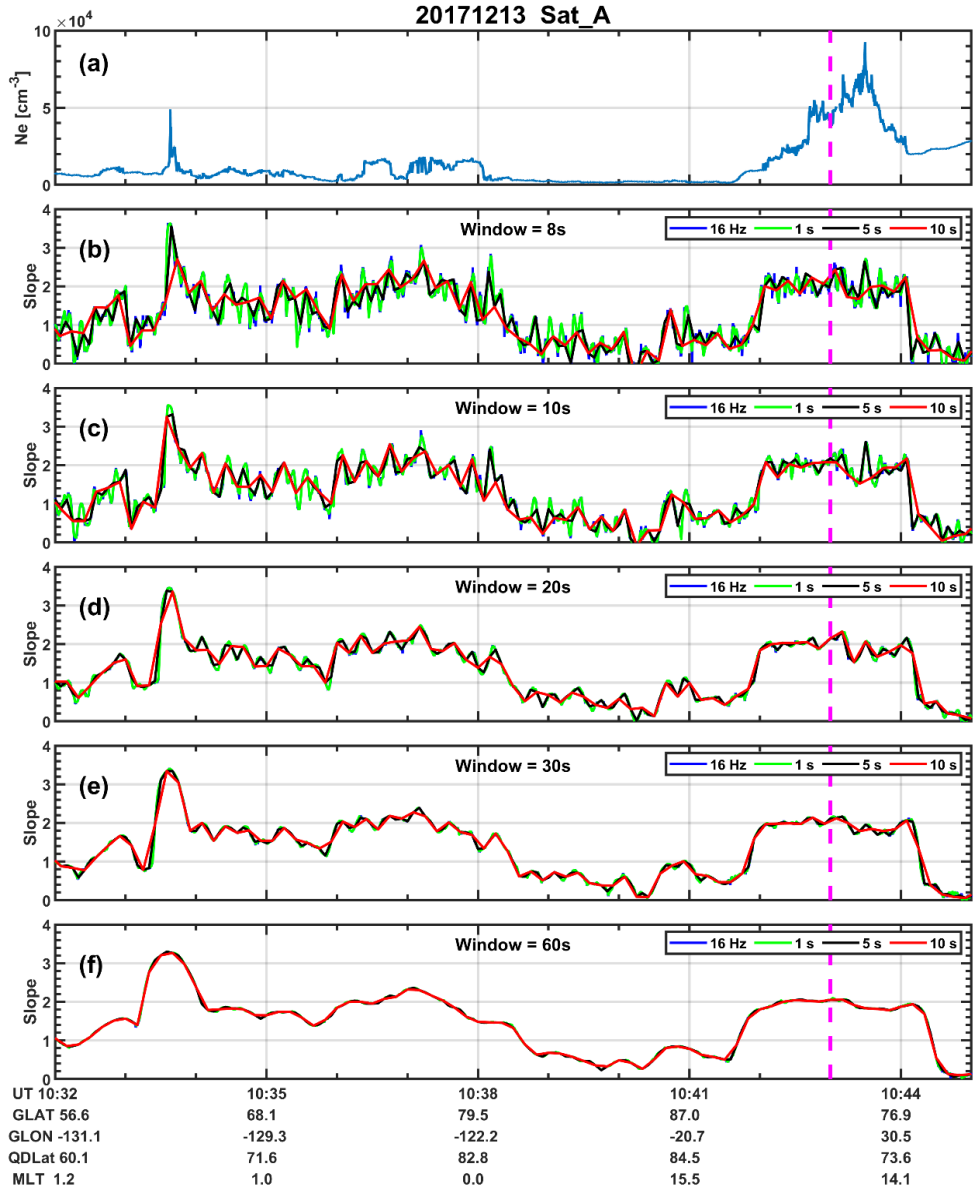

Figure 4. (a) The time series of the FP electron density. The spectral slopes calculated using window sizes of 8 s (b), 10 s (c), 20 s (d), 30 s (e), 60 s (f). Different colors are calculated using different time steps. The vertical dashed line presented the time of Figure 3.

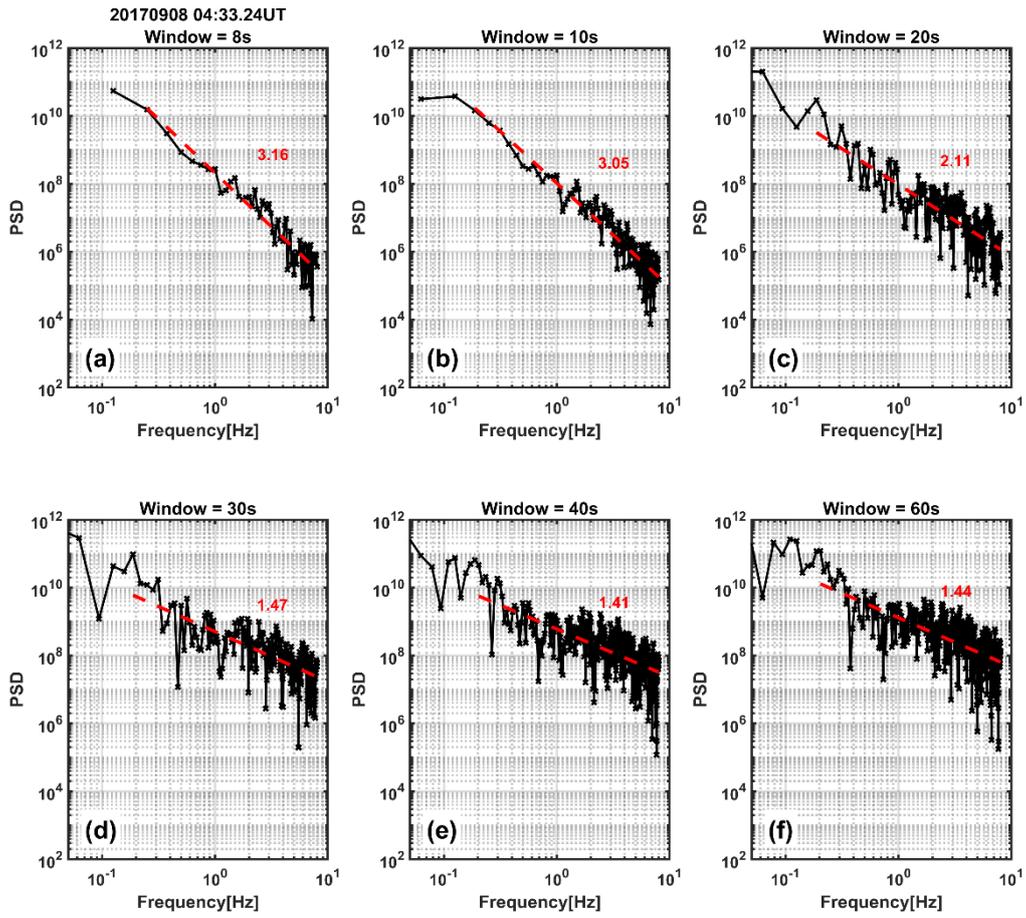

Figure 5. The same as Figure 3, but for equatorial plasma bubbles.

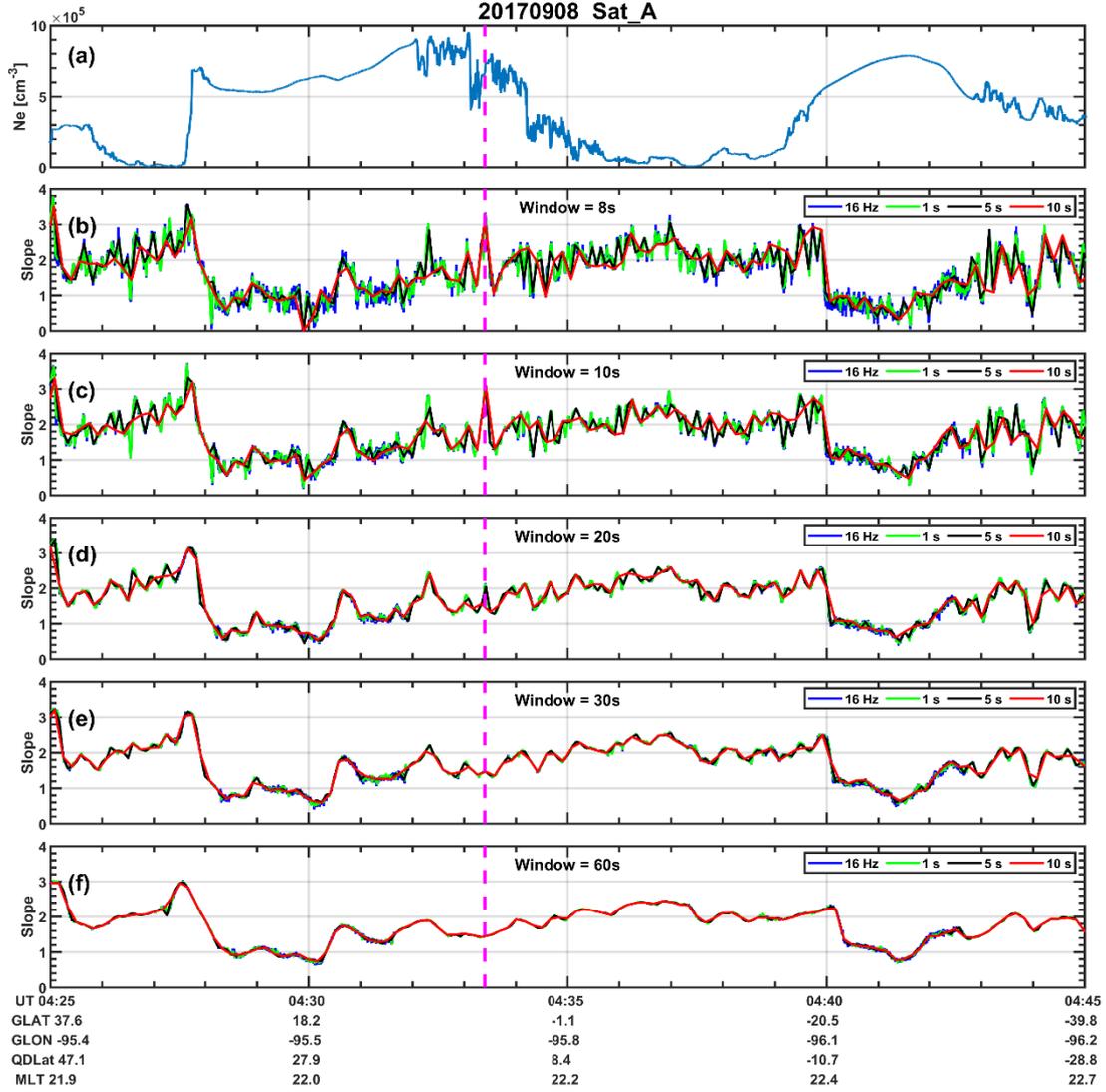

Figure 6. The same as Figure 4, but for equatorial latitudes. The vertical dashed line presented the time of Figure 5.

## 3. Statistical results

In this section, we show some statistics of the MUSIC dataset that is derived from the FP plasma density data. Figure 7 shows the long-term variations over the years 2014 to 2025 of electron density, $\nabla Ne@5\mathrm{km}$, $RODI1s$, and the percentage occurrence of spectral slope $p > 2$ from Swarm A, respectively. The statistical results for Swarm B and C can be found in the supplementary material. Note that Swarm C (with a slightly different longitudinal difference of 1.4°) is at the same altitude and almost the same local time (LT) as Swarm A, while Swarm B is at a higher altitude and different LT. In Figure 7, the data are divided into bins of 1 day and 2 degrees in the argument of quasi-dipole latitude. The availability of the FP plasma density data is determined by the operational schedule of the TII. Specifically, the 16 Hz plasma density data provided by the FP is only available when the TII is not in operation. The TII does not operate 100% of the time due to planned operational constraints and known instrument complexities (Knudsen et al., 2017; Burchill & Knudsen, 2022).

The FP plasma density data from Swarm A are available from late 2014 and remain almost continuously until late 2019. From late 2019 to mid-2023, the data are available only sporadically; however, after mid-2023, the availability of FP data has increased significantly. Note that the data are daily averages. This implies that, while FP data may not be available for the entire 24-hour period, they may still be available for specific orbits throughout that day.

To guide the analysis, Figure 7a shows F10.7, its moving average over 81 days (F10.7$_{81}$), and the daily averaged Kp index. The 81-day average is centered at the timestamp, i.e., using data from 40 days before and after the current day and the current day. As the LT of the ascending and descending orbits migrates slowly, we also overplot the LT of the ascending and descending orbits of Swarm A in solid and dashed red lines, respectively. Indeed, a key aspect of the Swarm orbits is that they drift in Local Time (LT). Consequently, it takes approximately 133 days for Swarm A and C, and 141 days for Swarm B, to cover all LT sectors (Wood et al., 2022).

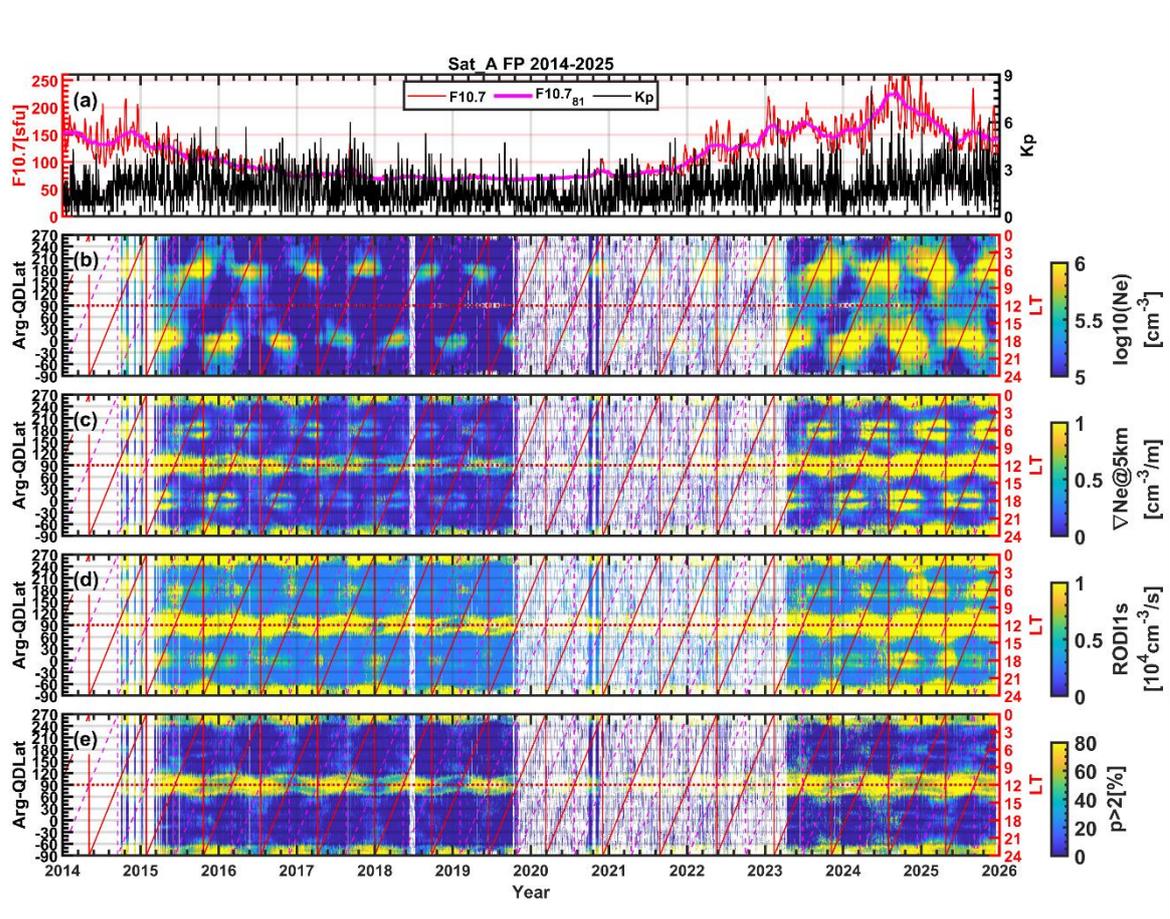

Figure 7. (a) The F10.7 index and the daily averaged Kp index. F10.7$_{81}$ is the moving average of F10.7 in a running window of 81 days. The long-term variations of electron density (b), density gradient at 5 km (c), *RODI1s* (d), and the occurrence of spectral slope *p* > 2 (e). The data are binned in 1 day and 2 degrees in the argument of quasi-dipole latitude. The argument latitude is defined as from -90° in the South Pole, 0° in the equator, and 90° in the North Pole for the ascending orbit, and it is 180° in the equator, 270° in the south pole for the descending orbit. The local time (LT) of the ascending and descending orbits are shown as solid and dashed red lines, respectively.

The general features of the ionosphere can be seen in Figure 7b, i.e., 1) higher density at low latitudes (around magnetic equator), which are enhanced between about 12-21 local time (LT); 2) higher density during solar maxima, 2015-2016, and 2023-2025. The density gradient at 5 km and *RODI1s* in Figures 7c-7d can be used to characterize plasma structures and irregularities, and irregularities are most prominent at low and high latitudes, in agreement also with what highlighted in Jin et al. (2020). At low latitudes, $\nabla Ne@5km$ is enhanced from about 12 LT and it becomes more enhanced after 18 LT. The enhanced values of $\nabla Ne@5km$ on the dayside (12-18 LT) are related to Equatorial Ionization Anomaly (EIA), which is associated with significant horizonal density gradients at a fixed altitude. The more enhanced $\nabla Ne@5km$ after sunset comes from equatorial plasma bubbles. At low latitudes, it is more appropriate to use *RODI1s* to investigate ionospheric irregularities (Jin et al., 2020). *RODI1s* is clearly enhanced from post-sunset hours (after 18 LT) for both ascending and descending orbits and they are mostly related with the occurrence of equatorial plasma bubbles able to reach the satellite altitude (see e.g., Spogli et al., 2023). At high latitudes, enhanced irregularities can be found above 60º QDLat for both hemispheres. Figure 7e shows the occurrence percentage of spectral slope *p* > 2 as an indication of the steepening of plasma density spectra (Ivarsen et al., 2021b). The occurrence of *p* > 2 is more enhanced at high latitudes because there are persistent high-latitude ionospheric irregularities which are enhanced at all scales, and this makes the calculation of the spectral slope from PSD more sensible. On the other hand, at mid- and low latitudes, the ionosphere is often smooth and lack of plasma irregularities. In this case, the calculation of spectral slope only makes mathematical sense. Though the spectral slope is often low at mid- and low latitudes, the occurrence of steep spectral slope (*p* > 2) is indeed visible when equatorial plasma irregularities exist, i.e., in regions of enhanced density gradients at 5 km and where *RODI1s* are enhanced. Another interesting feature of *p* > 2 is the seasonal variations at high latitudes. Near the polar regions, enhanced occurrences of *p* > 2 are observed near local summers and the occurrence minimizes near local winters for both hemispheres. This effect can be explained by the seasonal variations of the E region conductance in the polar regions (Ivarsen et al., 2019). The enhanced E region conductance in the summer hemisphere dampens small-scale ionospheric irregularities, and thus steepens the spectral slope of the PSD (Ivarsen et al., 2021b).

Figure 8 shows the statistical distribution of electron density, $\nabla Ne@5km$, *RODI1s*, and the occurrence of spectral slope *p* > 2 in QDLat and magnetic local time (MLT). The electron density is enhanced between 9-22 MLT and between ±25º QDLat. There is a region of very low density near the magnetic equator between 4-6 MLT. The main ionospheric trough of low ionospheric electron density is also visible at roughly 60º QDLat between 15 MLT and 8 MLT, and it is clearer (even lower electron density) in the morning side of the northern hemisphere. At high latitudes, the horizontal density gradient $\nabla Ne@5km$ is generally enhanced poleward of ±60º QDLat. The equatorward boundary of enhanced $\nabla Ne@5km$ shifts poleward toward 12 MLT. This is consistent with the MLT variations of the auroral oval. At low latitudes, enhanced $\nabla Ne@5km$ can be observed with EIA during 9-18 MLT. From sunset to post-midnight, enhanced $\nabla Ne@5km$ is associated with equatorial plasma bubbles. *RODI1s*, on the other hand, is only enhanced after 19.5 MLT till post-midnight. This is associated with plasma bubbles (Spogli et al., 2023), while the EIA is associated with low *RODI1s* activity. The occurrence of *p* > 2 is generally high at high latitudes and it seems to relate to $\nabla Ne@5km$ and *RODI1s*.

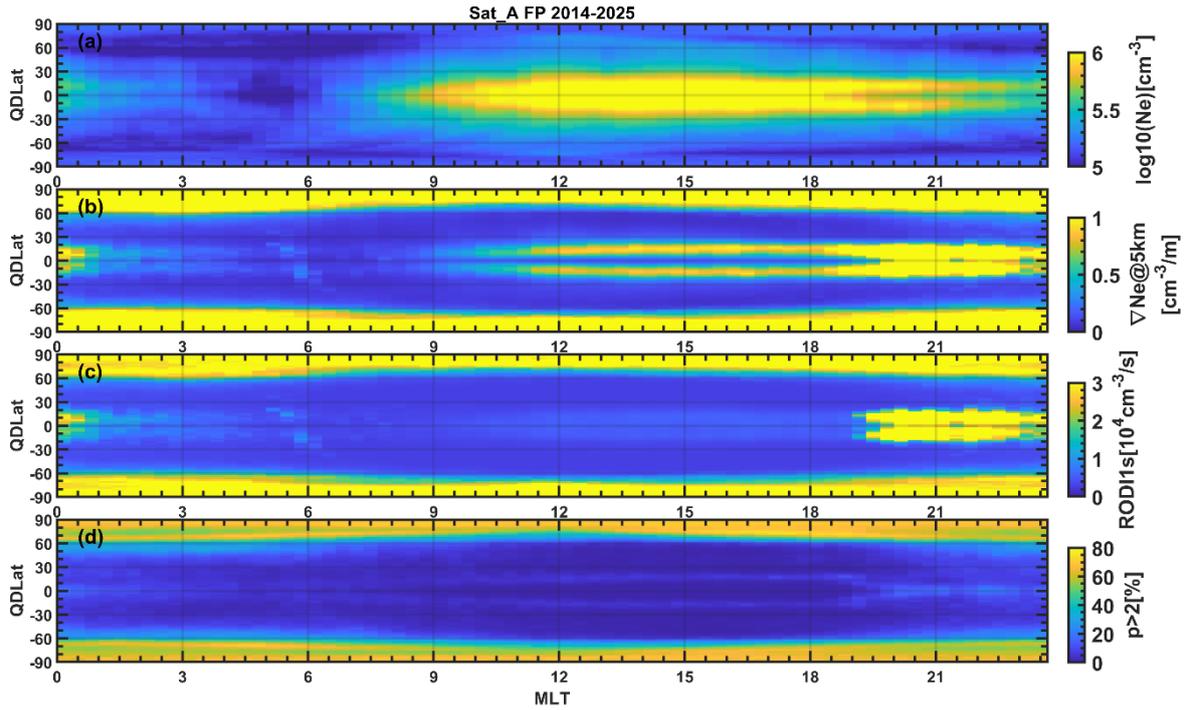

Figure 8. The distribution of (a) electron density, (b) electron density gradient at 5 km, (c) *RODI1s*, and (d) spectral slope as a function of quasi-dipole latitude and MLT.

Figure 9 shows the global maps of electron density, ∇*Ne*@5km, *RODI1s*, and the occurrence of spectral slope *p* > 2. The region of high electron density is roughly confined between ±20º QDLat. Figures 9b-9c show enhanced equatorial irregularities characterized by ∇*Ne*@5km and *RODI1s* mostly in regions from Africa to South America. Equatorial irregularities are not enhanced right above the magnetic equator but between 0º and ±20º QDLat. While these parameters also exhibit enhancement in other longitudinal sectors, their intensity is notably lower. This pronounced exacerbation of ionospheric irregularities and density gradients over the South American and Atlantic sectors is consistent with previous statistical analyses (see e.g., Yizengaw & Groves, 2018; Aa et al., 2020). These studies attribute the maximum occurrence rate of post-sunset equatorial plasma irregularities in this region to the weak geomagnetic field of the South Atlantic Anomaly region and the optimal alignment of magnetic declination with the solar terminator. At high latitudes, ∇*Ne*@5km and *RODI1s* are enhanced poleward of ±65º QDLat for both hemispheres. It is clear that ionospheric irregularities are controlled by magnetic coordinates at both low and high latitudes. The occurrence of *p* > 2 are very high (>60%) for high latitudes (poleward of ±65º QDLat), while it is very low (<10%) at low latitudes. In the literature, steep spectral slopes are frequently observed in the equatorial ionosphere (Aol et al., 2022).We will demonstrate later that this discrepancy can be ascribed to the relatively lower occurrence frequency of low-latitude irregularities, which are primarily restricted to the post-sunset period, as compared to high latitudes.

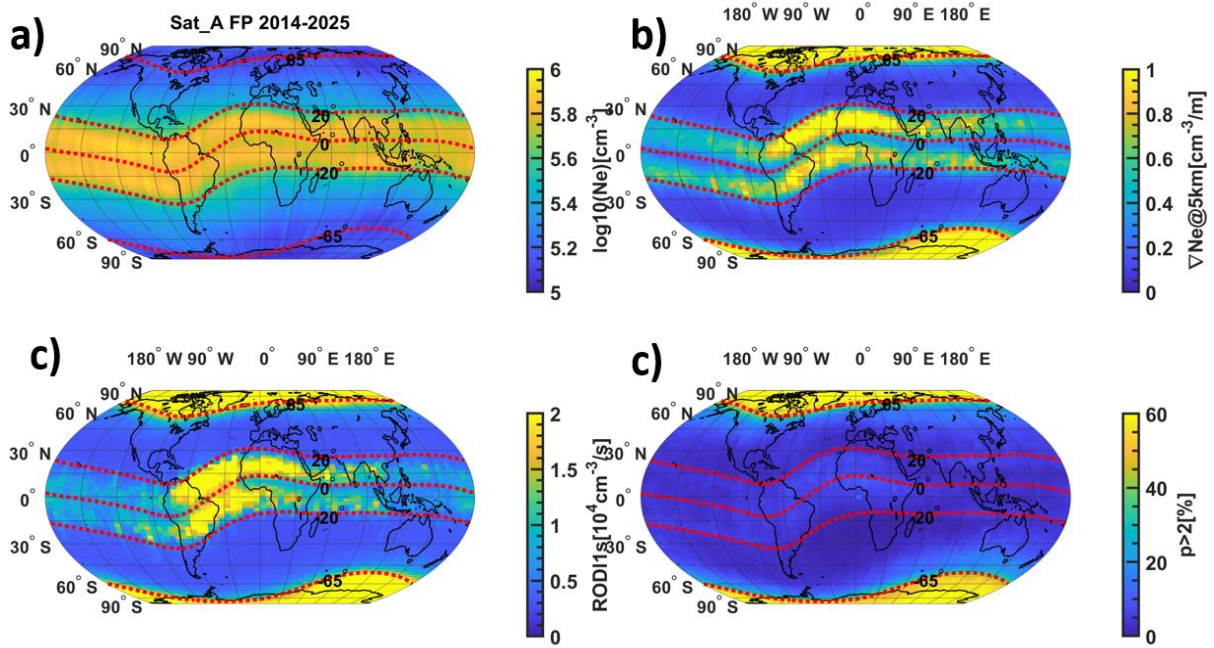

Figure 9. The distribution of (a) electron density, (b) electron density gradient at 5 km, (c) *RODI1s*, and (d) occurrence percentage of spectral slopes *p* > 2 in geographic coordinates. In each map, the magnetic latitudes of 0º, ±20º, and ±65º are plotted as red dotted lines.

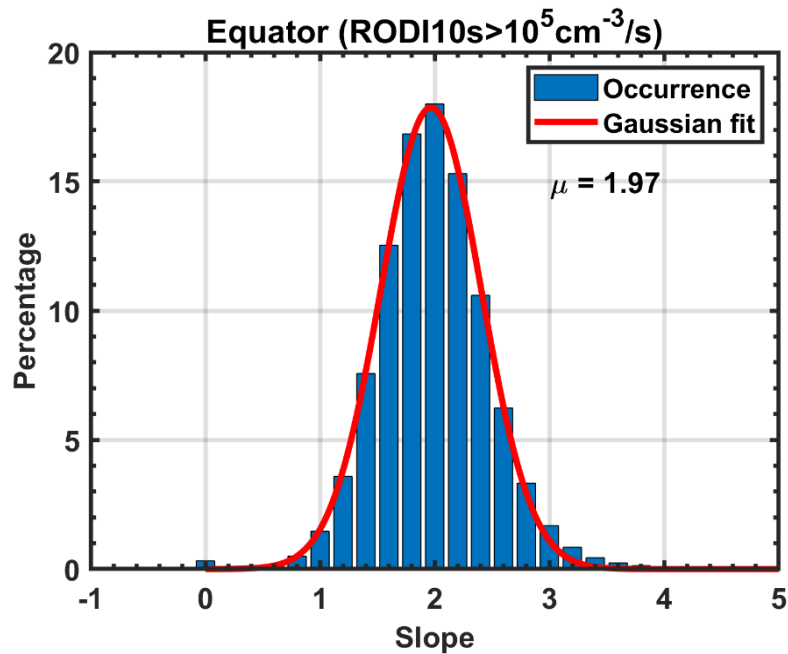

Figure 10. The histogram of spectral slope in the equatorial regions (equatorward of 30º QDLat) for *RODI10s* > 1x10$^5$ cm$^{-3}$/s. The red line is the Gaussian fit of the histogram with expected value μ = 1.97.

Figure 10 shows the histogram of spectral slopes in the equatorial regions (equatorward of 30º QDLat) during 2014-2025, featured by the presence of plasma irregularities The presence of irregularities is defined using the *RODI10s* index, with a threshold of *RODI10s* > 10$^5$ cm$^{-3}$/s as an indication of significant plasma irregularities. We make use of *RODI10s* because its 10-s window is the same as that used to calculate the PSD and the resulting spectral slope. The Gaussian fit (red) of the histogram shows an expected value of 1.97. In other words, if we calculate the occurrence rate of $p > 2$ for the condition of significant equatorial irregularities, it is approximately 50%. This value is considerably higher than the low occurrence of < 10% in Figures 8d and 9d. By using a similar method, Aol et al. (2022) presented a statistical distribution of the spectral slope during the time period 2014-2018 and they found a peak occurrence at $p = 2.5$. Despite the similarity with our result, the difference may be attributed to slightly different time windows (10 s in this study versus 8 s in Aol et al. (2022)) for calculating the PSD, the frequency range for fitting the spectral slope (0.2-8 Hz in this study versus 0.7-8 Hz in Aol et al. (2022)), different definitions of plasma irregularities etc.

### 4. Summary and conclusion

In this paper, we develop a new high-resolution data product for characterizing ionospheric structures and irregularities using the 16 Hz FP plasma density data from the Swarm mission. Due to the high resolution, the data product is able to capture electron density variations down to sub-kilometer scales, providing a significant advantage compared to the IPIR product, which relies on 2 Hz LP data. A primary constraint of the FP data is its intermittent availability, because the FP plasma density data are only derived when the bias of TII is below -2.5 V. For example, the FP data from Swarm A are only sporadically available between 2020-2023, while the data are more continuous during other time periods. In total, approximately 8 years of data from Swarm A have been processed, providing a meaningful dataset for robust long-term statistical analysis of ionospheric variability at the smallest available spatial scale. Furthermore, the increasing recognition of FP data as a valuable resource for investigating small-scale phenomena has led to ongoing efforts to maximize its duty cycle (see e.g., Smirnov et al., 2024).

Based on the data spanning late 2014 to 2025, we present global statistics of the electron density, $\nabla Ne@5km$, *RODI1s* and the occurrence percentage of spectral slope $p > 2$. The provided statistics show regions of enhanced density gradients and irregularities in both low and high latitudes. The occurrence of $p > 2$ is more pronounced at high latitudes. In addition, there are seasonal variations in the occurrence of $p > 2$ at high latitudes, i.e., the occurrence maximizes during local summer and minimizes during local winter. This effect can be explained by the solar produced E region conductance that maximizes during local summer. The lower occurrence of $p > 2$ at low latitudes is due to the relatively lower occurrence of ionospheric irregularities at low latitudes as compared to high latitudes. When we confine the data to *RODI10s* > $1\times10^5$ cm$^{-3}$/s, the occurrence of $p > 2$ reaches 50%. This implies that the spectral slope $p$ is more meaningful when enough ionospheric plasma structures and irregularities are present.

The validity of absolute FP density values remains a point of discussion. For example, it is known that the absolute values of the FP data are generally higher than the LP plasma density data (Buchert, 2017; Catapano et al., 2022). There have been a few studies to validate the FP data. For example, Xiong et al. (2022) systematically compared the FP and LP plasma density data and found differences depending on the solar flux, local time, and season. However, the LP plasma density data itself is dependent on the solar flux (possibly due to the chemical composition variations in the ionospheric plasma), while such dependence cannot be found in the FP data (Xiong et al., 2022). In addition, the LP plasma density data was found to be lower than those obtained from Incoherent Scatter Radar (Lomidze et al., 2018). Therefore, there is no simple and straightforward way to validate and calibrate the absolute values of the FP data. A recent attempt to calibrate LP data using FP measurements has been introduced by Smirnov et al. (2024). This approach utilizes neural networks and is aimed at reducing ion density overestimations in the nightside ionosphere at mid-latitudes.

Besides, the FP data have been successfully used to investigate ionospheric irregularities at both high and low latitudes (Ivarsen et al., 2021a; Ivarsen et al., 2021b; Aol et al., 2022; Zheng et al., 2025). The relative variations derived from the FP data should be sufficient to study small-scale ionospheric phenomena, with some usefulness also for L-band scintillation studies and modelling (Aol et al., 2020; Zheng et al., 2025). Given the systematic differences in the absolute values of FP and LP data, similar differences should be present in the derived parameters of the MUSIC dataset ($\nabla Ne@20km$, $\nabla Ne@50km$, $\nabla Ne@100km$, *RODI10s*, and *RODI20s*) as well.

The MUSIC data that contains a set of parameters to characterize multi-scale ionospheric structures and irregularities along Swarm orbit are open to the wider community and are available in CDF format: http://tid.uio.no/plasma/swarm/IPIR_cdf/products_FP/. Given the high resolution, this new data product can provide new possibility for the amplitudes, variability, scale sizes and global extent of multi-scale ionospheric structures and irregularities. We note that the smallest spatial scale is close to the Fresnel's scale (~400 m) for GNSS scintillations. This makes it particularly useful for GNSS users and different groups working on the development of models and services for ionospheric and space weather effects on the GNSS signals. In addition, the scientific communities that are interested in the magnetosphere-ionosphere-thermosphere coupling and near-Earth space environment will also benefit from the MUSIC dataset.

**Acknowledgments**

This research is a part of the 4DSpace Strategic Research Initiative at the University of Oslo. The authors wish to thank Rayan Iman, Eelco Doornbos, Kasper van dam and Elisabetta Iorfida for useful discussions.

**Funding**

This work is funded under the European Space Agency (ESA) Contract 4000143413/23/ I-EB (Swarm VIP Dynamics) within the ESA Solid and Magnetic Science Cluster - 4D Ionosphere framework. YJ, DK and WJM acknowledge funding from the European Research Council (ERC) under the European Union's Horizon 2020 research and innovation programme (ERC Consolidator Grant 866357, POLAR-4DSpace).

**Data Availability Statement**

The Swarm data can be obtained through the official Swarm website (https://earth.esa.int/eogateway/missions/swarm/data) and for the FP plasma density data go to https://swarm-diss.eo.esa.int/#swarm%2FAdvanced%2FPlasma_Data%2F16_Hz_Faceplate_plasma_density%2FSat_B. The processed MUSIC data in this paper is now available in CDF format at: http://tid.uio.no/plasma/swarm/IPIR_cdf/products_FP/. The F10.7 and Kp indices were obtained from the NASA OMNIWeb service (https://omniweb.gsfc.nasa.gov/form/dx1.html).

**References**

Aa, E, Zou, S, Liu, S. 2020. Statistical Analysis of Equatorial Plasma Irregularities Retrieved From Swarm 2013–2019 Observations. *Journal of Geophysical Research: Space Physics, 125*(4), e2019JA027022. https://doi.org/10.1029/2019ja027022

Aarons, J. 1982. Global morphology of ionospheric scintillations. *Proceedings of the IEEE, 70*(4), 360-378. https://doi.org/10.1109/proc.1982.12314

Alfonsi, L, Spogli, L, De Franceschi, G, Romano, V, Aquino, M, et al. 2011. Bipolar climatology of GPS ionospheric scintillation at solar minimum. *Radio Science, 46*(3). https://doi.org/10.1029/2010rs004571

Aol, S, Buchert, S, Jurua, E, Milla, M. 2020. Simultaneous ground-based and in situ Swarm observations of equatorial F-region irregularities over Jicamarca. *Annales Geophysicae, 38*(5), 1063-1080. https://doi.org/10.5194/angeo-38-1063-2020

Aol, S, Buchert, S, Jurua, E, Sorriso-Valvo, L. 2022. Spectral properties of sub-kilometer-scale equatorial irregularities as seen by the Swarm satellites. *Advances in Space Research, 72*(3), 741-752. https://doi.org/10.1016/j.asr.2022.07.059

Balan, N, Liu, L, Le, H. 2018. A brief review of equatorial ionization anomaly and ionospheric irregularities. *Earth and Planetary Physics, 2*(4), 257-275. https://doi.org/10.26464/epp2018025

Basu, S, Mackenzie, E, Basu, S. 1988. Ionospheric Constraints on VHF UHF Communications Links during Solar Maximum and Minimum Periods. *Radio Science, 23*(3), 363-378. https://doi.org/10.1029/RS023i003p00363

Bhattacharyya, A. 2022. Equatorial plasma bubbles: A review. *Atmosphere, 13*(10), 1637. https://doi.org/10.3390/atmos13101637

Buchert, S. (2017). Extended EFI LP Data FP Release Notes. Retrieved from https://swarmhandbook.earth.esa.int/catalogue/sw_efix_lp_fp

Burchill, JK, Knudsen, DJ. 2022. Swarm thermal ion imager measurement performance. *Earth, Planets and Space, 74*(1), 181. https://doi.org/10.1186/s40623-022-01736-w

Buschmann, LM, Clausen, L, Spicher, A, Ivarsen, MF, Miloch, WJ. 2024. Statistical studies of plasma structuring in the auroral ionosphere by the Swarm satellites. *Journal of Geophysical Research: Space Physics, 129*(2), e2023JA032097. https://doi.org/10.1029/2023JA032097

Catapano, F, Buchert, S, Qamili, E, Nilsson, T, Bouffard, J, et al. 2022. Swarm Langmuir probes' data quality validation and future improvements. *Geosci. Instrum. Method. Data Syst., 11*(1), 149-162. https://doi.org/10.5194/gi-11-149-2022

De Michelis, P, Consolini, G, Tozzi, R, Pignalberi, A, Pezzopane, M, et al. 2021. Ionospheric turbulence and the equatorial plasma density irregularities: Scaling features and RODI. *Remote Sensing, 13*(4), 759. https://doi.org/10.3390/rs13040759

Foster, JC, Coster, AJ, Erickson, PJ, Holt, JM, Lind, FD, et al. 2005. Multiradar observations of the polar tongue of ionization. *Journal of Geophysical Research-Space Physics, 110*(A9). https://doi.org/10.1029/2004ja010928

Friis-Christensen, E, Lühr, H, Hulot, G. 2006. Swarm: A constellation to study the Earth's magnetic field. *Earth, Planets and Space, 58*(4), 351-358. https://doi.org/10.1186/BF03351933

Fukao, S, Ozawa, Y, Yokoyama, T, Yamamoto, M, Tsunoda, RT. 2004. First observations of the spatial structure of F region 3-m-scale field-aligned irregularities with the Equatorial Atmosphere Radar in Indonesia. *Journal of Geophysical Research: Space Physics, 109*(A2). https://doi.org/10.1029/2003JA010096

Ivarsen, MF, Jin, Y, Spicher, A, Clausen, LB. 2019. Direct evidence for the dissipation of small‐scale ionospheric plasma structures by a conductive E region. *Journal of Geophysical Research: Space Physics, 124*(4), 2935-2942. https://doi.org/10.1029/2019JA026500

Ivarsen, MF, Jin, YQ, Spicher, A, Miloch, W, Clausen, LBN. 2021a. The Lifetimes of Plasma Structures at High Latitudes. *Journal of Geophysical Research-Space Physics, 126*(2). https://doi.org/10.1029/2020ja028117

Ivarsen, MF, St-Maurice, JP, Jin, YQ, Park, J, Miloch, W, et al. 2021b. Steepening Plasma Density Spectra in the Ionosphere: The Crucial Role Played by a Strong E-Region. *Journal of Geophysical Research-Space Physics, 126*(8). https://doi.org/10.1029/2021ja029401

Jin, Y, Kotova, D, Clausen, LB, Miloch, WJ. 2025. Significant plasma density depletion from high-to mid-latitude ionosphere during the super storm in May 2024. *Geophysical Research Letters, 52*(5), e2024GL113997. https://doi.org/10.1029/2024GL113997

Jin, Y, Kotova, D, Xiong, C, Brask, SM, Clausen, LBN, et al. 2022. Ionospheric Plasma IRregularities - IPIR - Data Product Based on Data From the Swarm Satellites. *Journal of Geophysical Research: Space Physics, 127*(4), e2021JA030183. https://doi.org/10.1029/2021JA030183

Jin, Y, Xiong, C. 2020. Interhemispheric Asymmetry of Large-Scale Electron Density Gradients in the Polar Cap Ionosphere: UT and Seasonal Variations. *Journal of Geophysical Research: Space Physics, 125*(2), e2019JA027601. https://doi.org/10.1029/2019ja027601

Jin, Y, Xiong, C, Clausen, L, Spicher, A, Kotova, D, et al. 2020. Ionospheric Plasma Irregularities Based on In Situ Measurements From the Swarm Satellites. *Journal of Geophysical Research: Space Physics, 125*(7), e2020JA028103. https://doi.org/10.1029/2020JA028103

Jin, YQ, Moen, JI, Miloch, WJ. 2014. GPS scintillation effects associated with polar cap patches and substorm auroral activity: direct comparison. *Journal of Space Weather and Space Climate, 4*(A23). https://doi.org/10.1051/swsc/2014019

Jin, YQ, Moen, JI, Miloch, WJ. 2015. On the collocation of the cusp aurora and the GPS phase scintillation: A statistical study. *Journal of Geophysical Research-Space Physics, 120*(10), 9176-9191. https://doi.org/10.1002/2015ja021449

Jin, YQ, Spicher, A, Xiong, C, Clausen, LBN, Kervalishvili, G, et al. 2019. Ionospheric Plasma Irregularities Characterized by the Swarm Satellites: Statistics at High Latitudes. *Journal of Geophysical Research-Space Physics, 124*(2), 1262-1282. http://doi.org/10.1029/2018ja026063

Kelley, MC, Vickrey, JF, Carlson, CW, Torbert, R. 1982. On the Origin and Spatial Extent of High-Latitude F-Region Irregularities. *Journal of Geophysical Research-Space Physics, 87*(Na6), 4469-4475. https://doi.org/10.1029/JA087iA06p04469

Kintner, PM, Ledvina, BM, De Paula, ER. 2007. GPS and ionospheric scintillations. *Space Weather-the International Journal of Research and Applications, 5*(9). https://doi.org/10.1029/2006sw000260

Kintner, PM, Seyler, CE. 1985. The status of observations and theory of high latitude ionospheric and magnetospheric plasma turbulence. *Space Science Reviews, 41*(1), 91-129. https://doi.org/10.1007/BF00241347

Knudsen, DJ, Burchill, JK, Buchert, SC, Eriksson, AI, Gill, R, et al. 2017. Thermal ion imagers and Langmuir probes in the Swarm electric field instruments. *Journal of Geophysical Research-Space Physics, 122*(2), 2655-2673. https://doi.org/10.1002/2016ja022571

Kotova, D, Jin, Y, Miloch, W. 2022. Interhemispheric variability of the electron density and derived parameters by the Swarm satellites during different solar activity. *Journal of Space Weather and Space Climate, 12*, 12. https://doi.org/10.1051/swsc/2022007

Kotova, D, Jin, Y, Spogli, L, Wood, AG, Urbar, J, et al. 2023. Electron density fluctuations from Swarm as a proxy for ground-based scintillation data: A statistical perspective. *Advances in Space Research, 72(12)*(12), 5399-5415. https://doi.org/10.1016/j.asr.2022.11.042

Lomidze, L, Knudsen, DJ, Burchill, J, Kouznetsov, A, Buchert, SC. 2018. Calibration and Validation of Swarm Plasma Densities and Electron Temperatures Using Ground-Based Radars and Satellite Radio Occultation Measurements. *Radio Science, 53*(1), 15-36. https://doi.org/10.1002/2017rs006415

Mitchell, CN, Alfonsi, L, De Franceschi, G, Lester, M, Romano, V, Wernik, AW. 2005. GPS TEC and scintillation measurements from the polar ionosphere during the October 2003 storm. *Geophysical Research Letters, 32*(12). https://doi.org/10.1029/2004gl021644

Moen, J, Carlson, HC, Oksavik, K, Nielsen, CP, Pryse, SE, et al. 2006. EISCAT observations of plasma patches at sub-auroral cusp latitudes. *Annales Geophysicae, 24*(9), 2363-2374. https://doi.org/10.5194/angeo-24-2363-2006

Moen, J, Carlson, HC, Rinne, Y, Skjaeveland, A. 2012. Multi-scale features of solar terrestrial coupling in the cusp ionosphere. *Journal of Atmospheric and Solar-Terrestrial Physics, 87-88*, 11-19. https://doi.org/10.1016/j.jastp.2011.07.002

Nishimura, Y, Deng, Y, Lyons, LR, McGranaghan, RM, Zettergren, MD. 2021. Multiscale Dynamics in the High-Latitude Ionosphere. In *Ionosphere Dynamics and Applications* (pp. 49-65) https://doi.org/10.1002/9781119815617.ch3

Ovodenko, VB, Klimenko, MV, Zakharenkova, IE, Oinats, AV, Kotova, DS, et al. 2020. Spatial and Temporal Evolution of Different-Scale Ionospheric Irregularities in Central and East Siberia During the 27–28 May 2017 Geomagnetic Storm. *Space Weather, 18*(6), e2019SW002378. https://doi.org/10.1029/2019SW002378

Smirnov, A, Shprits, Y, Lühr, H, Pignalberi, A, Xiong, C. 2024. Calibration of Swarm plasma densities overestimation using neural networks. *Space Weather, 22*(8), e2024SW003925. https://doi.org/10.1029/2024SW003925

Song, H, Park, J, Jin, Y, Otsuka, Y, Buchert, S, et al. 2023. Tandem Observations of Nighttime Mid-Latitude Topside Ionospheric Perturbations. *Space Weather, 21*(2), e2022SW003312. https://doi.org/10.1029/2022SW003312

Spogli, L, Alfonsi, L, Cesaroni, C. 2023. Stepping into an equatorial plasma bubble with a Swarm overfly. *Space Weather, 21*(5), e2022SW003331. https://doi.org/10.1029/2022SW003331

Spogli, L, Alfonsi, L, De Franceschi, G, Romano, V, Aquino, MHO, Dodson, A. 2009. Climatology of GPS ionospheric scintillations over high and mid-latitude European regions. *Annales Geophysicae, 27*(9), 3429-3437. https://doi.org/10.5194/angeo-27-3429-2009

Spogli, L, Jin, Y, Urbář, J, Wood, AG, Donegan-Lawley, EE, et al. 2024. Statistical models of the variability of plasma in the topside ionosphere: 2. Performance assessment. *Journal of Space Weather and Space Climate, 14*, 4. https://doi.org/10.1051/swsc/2024003

Tsunoda, RT. 1988. High-Latitude F-Region Irregularities - a Review and Synthesis. *Reviews of Geophysics, 26*(4), 719-760. https://doi.org/10.1029/RG026i004p00719

Urbar, J, Spogli, L, Cicone, A, Clausen, LB, Jin, Y, et al. 2022. Multi-scale response of the high-latitude topside ionosphere to geospace forcing. *Advances in Space Research, 72*(12), 5490-5502. https://doi.org/10.1016/j.asr.2022.06.045

Wang, Y, Zhang, Q-H, Jayachandran, PT, Lockwood, M, Zhang, S-R, et al. 2016. A comparison between large-scale irregularities and scintillations in the polar ionosphere. *Geophysical Research Letters, 43*(10), 4790-4798. https://doi.org/10.1002/2016GL069230

Wood, AG, Alfonsi, L, Clausen, LBN, Jin, Y, Spogli, L, et al. 2022. Variability of Ionospheric Plasma: Results from the ESA Swarm Mission. *Space Science Reviews, 218*(6), 52. https://doi.org/10.1007/s11214-022-00916-0

Wood, AG, Donegan-Lawley, EE, Clausen, LBN, Spogli, L, Urbář, J, et al. 2024. Statistical models of the variability of plasma in the topside ionosphere: 1. Development and optimisation. *J. Space Weather Space Clim., 14*, 7. https://doi.org/10.1051/swsc/2024003

Xiong, C, Jiang, H, Yan, R, Lühr, H, Stolle, C, et al. 2022. Solar Flux Influence on the In-Situ Plasma Density at Topside Ionosphere Measured by Swarm Satellites. *Journal of Geophysical Research: Space Physics, 127*(5), e2022JA030275. https://doi.org/10.1029/2022JA030275

Xiong, C, Park, J, Luehr, H, Stolle, C, Ma, SY. 2010. Comparing plasma bubble occurrence rates at CHAMP and GRACE altitudes during high and low solar activity. *Annales Geophysicae, 28*(9), 1647-1658. https://doi.org/10.5194/angeo-28-1647-2010

Yeh, KC, Liu, CH. 1982. Radiowave Scintillations in the Ionosphere. *Proceedings of the Ieee, 70*(4), 324-360. https://doi.org/10.1109/PROC.1982.12313

Yizengaw, E, Groves, KM. 2018. Longitudinal and seasonal variability of equatorial ionospheric irregularities and electrodynamics. *Space Weather, 16*(8), 946-968. https://doi.org/10.1029/2018SW001980

Zakharenkova, I, Astafyeva, E. 2015. Topside ionospheric irregularities as seen from multisatellite observations. *Journal of Geophysical Research-Space Physics, 120*(1), 807-824. https://doi.org/10.1002/2014ja020330

Zhang, QH, Ma, YZ, Jayachandran, PT, Moen, J, Lockwood, M, et al. 2017. Polar cap hot patches: Enhanced density structures different from the classical patches in the ionosphere. *Geophysical Research Letters, 44*(16), 8159-8167. https://doi.org/10.1002/2017gl073439

Zheng, Y, Xiong, C, Spogli, L, Iman, R, Alfonsi, L. 2025. Global ionospheric scintillation estimation based on phase screen modeling from one-dimensional satellite data. *Space Weather, 23*(1), e2024SW004103. https://doi.org/10.1029/2024SW004103